%% file: main.tex
\documentclass[runningheads]{llncs}
\usepackage{graphicx}
\usepackage[margin=3cm]{geometry}
\usepackage[colorlinks=false,hidelinks]{hyperref}
\usepackage{amsmath,amssymb}
\usepackage{cleveref}
\usepackage{braket}
\usepackage{tikz}
\usetikzlibrary{quantikz2}
\usepackage[customcolors,shade]{hf-tikz}
\usepackage{caption}
\usepackage{subcaption}
\usepackage{todonotes}

\newcommand{\ketbra}[2]{\ket{#1}\!\!\bra{#2}}
\newcommand{\ketbras}[1]{\ketbra{#1}{#1}}

\begin{document}
\title{Circuit Cutting with Non-Maximally Entangled States}
%
%
\author{Marvin Bechtold \and
Johanna Barzen \and
Frank Leymann \and
Alexander Mandl
}
\authorrunning{M. Bechtold et al.}
%
\institute{Institute of Architecture of Application Systems \\
University of Stuttgart\\ 
Universitätsstraße 38, 70569 Stuttgart, Germany\\
\email{\{bechtold,barzen,leymann,mandl\}@iaas.uni-stuttgart.de}}
\maketitle              
\begin{abstract}
\input{content/0_abstract}

\end{abstract}

\input{content/1_introduction}

\input{content/2_preliminaries}

\input{content/3_main_result}

\input{content/4_numerical_experiments}

\input{content/5_related_work}

\input{content/6_discussion_conclusion}

\section*{Acknowledgement}
This work was partially funded by the projects \textit{PlanQK} (01MK20005N) and \textit{EniQmA} (01MQ22007B) of the \textit{Federal Ministry for Economic Affairs and Climate Action}.




%
%
%
\bibliographystyle{splncs04}
\bibliography{bibliography}
\end{document}

%% file: content/0_abstract.tex
Distributed quantum computing combines the computational power of multiple devices to overcome the limitations of individual devices.
Circuit cutting techniques enable the distribution of quantum computations through classical communication.
These techniques involve partitioning a quantum circuit into smaller subcircuits, each containing fewer qubits.
The original circuit's outcome can be replicated by executing these subcircuits on separate devices and combining their results.
However, the number of shots required to achieve a fixed result accuracy with circuit cutting grows exponentially with the number of cuts, posing significant costs.
In contrast, quantum teleportation allows the distribution of quantum computations without an exponential increase in shots.
Nevertheless, each teleportation procedure requires a pre-shared pair of maximally entangled qubits for transmitting a quantum state, and non-maximally entangled qubits cannot be used for this purpose.
To address this, we propose a novel circuit cutting technique that leverages non-maximally entangled qubit pairs, effectively reducing the cost associated with circuit cutting. 
By considering the degree of entanglement in the pre-shared qubit pairs, our technique provides a continuum between existing circuit cutting methods and quantum teleportation, adjusting the cost of circuit cutting accordingly. 

%% file: content/1_introduction.tex
\section{Introduction}
Quantum computing holds immense potential for solving problems intractable for classical computers~\cite{Cao2018,Cao2019}. 
However, contemporary quantum devices face various constraints, including a limited number of available qubits, and scaling up to larger devices remains a considerable challenge~\cite{Preskill2018}. 
One promising approach to address this limitation involves utilizing multiple smaller quantum devices that can exchange either classical~\cite{Avron2021,Dunjko2018} or, in the near future, quantum information~\cite{Cuomo2020,Khait2023}.
These modular systems enable distributed quantum computation, offering a pathway to scalability~\cite{Bravyi2022}. 
Moreover, as quantum devices mature, the integration of multiple quantum devices into the edge-cloud compute continuum is anticipated~\cite{Furutanpey2023_QuantumEdge}, further emphasizing the importance of distributed quantum computation.
\looseness=-1

Circuit cutting is a technique used to distribute the computation of a quantum circuit over multiple devices that can exchange only classical information~\cite{Bravyi2016,Brenner2023,Mitarai2021,Peng2019,Piveteau2022}.
This technique involves decomposing a large quantum circuit into smaller subcircuits by cutting it, allowing for their execution on multiple smaller quantum devices.
The subcircuit results can then be recombined through classical postprocessing to reproduce the outcome of the original quantum circuit. 
Although circuit cutting facilitates the execution of large quantum circuits on smaller devices, it introduces an overhead in required shots that scales exponentially with the number of cuts to achieve a fixed result accuracy.
This limits the applicability of circuit cutting to specific circuits amenable to cutting. 
Therefore, minimizing the overhead of each cut is crucial to extend the applicability of circuit cutting.

Quantum teleportation is an alternative technique that is essential for distributing quantum computations across multiple devices~\cite{Cuomo2020}.
In contrast to circuit cutting, it eliminates the need for additional shots when executing a quantum circuit across multiple devices by leveraging maximally entangled states. 
Based on a pre-shared maximally entangled state, quantum teleportation allows the transmission of quantum information between quantum devices by exchanging only two classical bits~\cite{Bennett1993}. 
Although generating these entangled qubit pairs between multiple commercial quantum devices is currently unattainable, it is anticipated to become feasible in the near future~\cite{Bravyi2022}.

The techniques of quantum teleportation and circuit cutting represent two extremes in terms of the amount of entanglement utilized. 
Quantum teleportation relies on maximally entangled states, while circuit cutting operates without any entanglement. 
However, neither of these techniques allows utilizing states with intermediate levels of entanglement, commonly referred to as \textit{non-maximally entangled~(NME)} states.
This study explores the potential of leveraging NME states with varying degrees of entanglement. 
The primary objective is to investigate how these states can be employed to reduce the overhead associated with circuit cutting.

This work is structured as follows:
\Cref{sec:preliminaries} establishes the preliminaries required for our novel approach utilizing NME states for circuit cutting in \Cref{sec:main-section}.
Subsequently, in \Cref{sec:num_experiments}, we present numerical experiments conducted with our approach.
\Cref{seq:related_work} provides a summary of the related work and \Cref{sec:conclusion} concludes the paper.

%% file: content/2_preliminaries.tex
\section{Preliminaries}\label{sec:preliminaries}
This section gives a concise introduction to wire cutting, a circuit cutting technique that is used in this work.
Furthermore, this section describes the NME states utilized in our modified wire cutting procedure, which is elaborated upon in \Cref{sec:main-section}.
Lastly, the quantum teleportation protocol is briefly discussed to facilitate a comprehensive understanding of the modifications made.

\subsection{Wire cutting}
Initially proposed by Peng et al.~\cite{Peng2019}, wire cutting is a circuit cutting technique that decomposes a wire into a linear combination of measurements and qubit initializations~\cite{Harada2023,Pednault2023}. 
Each measurement and subsequent initialization can be represented by a channel $\mathcal{E}$ of the form
\begin{align}
\mathcal{E}(\bullet) := \sum_{i} a_{i} \mathrm{Tr}[E_{i}(\bullet)] \rho_{i},
\end{align}
to which it is referred to as a \textit{measure-and-prepare channel}.
Herein,  $a_{i} = \pm 1$ and the operators ${E_{i}}$ form a positive operator-valued measure satisfying $\sum_{i} E_i = I$, with $E_{i}$ being positive semi-definite Hermitian matrices.
The one-qubit density operators $\rho_{i}$ are assumed to be efficiently preparable in quantum circuits.
The measurements and qubit initializations are designed to collectively produce the identity channel, which is defined as $\operatorname{Id}(\rho) = I\rho I$, where $\rho$ is a density operator.
Therefore, the wire cutting procedure decomposes the identity channel $\operatorname{Id}$ into $m$ measure-and-prepare channels:
\begin{align}\label{eq:id_decomposition}
\operatorname{Id}(\bullet) = \sum_{i=1}^{m} c_{i} \mathcal{E}_{i}(\bullet)
\end{align}
where $c_{i}$ are real coefficients that may take negative values, subject to the constraint $\sum_{i=1}^{m} c_{i} = 1$. 
These types of decompositions, which incorporate negative coefficients, are commonly referred to as a \textit{quasi-probabilistic decomposition}~\cite{Brenner2023,Piveteau2022}.

Using this decomposition of the identity channel, the expectation value of a state described by density operator $\rho$  with respect to observable $O$ can be written as
\begin{align}\label{eq:expectation}
\mathrm{Tr}[O\operatorname{Id}(\rho)] = \sum_{i=1}^{m} p_{i} \mathrm{Tr}[O\mathcal{E}_{i}(\rho)] \operatorname{sign}(c_i) \kappa
\end{align}
where $\kappa := \sum_i |c_i|$ and $p_i := |c_{i}|/\kappa$.
This enables the computation of the expectation value using a Monte Carlo approach~\cite{Brenner2023}: for each shot, an operation $\mathcal{E}_{i}$ is randomly selected with probability $p_{i}$, then the small subcircuits associated with operation $\mathcal{E}_{i}$ are executed, and the outcome is weighted by $\operatorname{sign}(c_i) \kappa$. 
The constructed estimator preserves the expectation value  but increases the variance by $\kappa$. 
To achieve a fixed statistical accuracy, estimating the expectation value within the error $\epsilon$ requires an additional $\mathcal{O}(\kappa^2/\epsilon^2)$ shots, commonly referred to as \textit{sampling overhead}~\cite{Temme2017}.
Minimizing the sampling overhead is highly desirable, and thus, the search for a decomposition that minimizes the value of $\kappa$ is of great importance. 
Notably, the minimum sampling overhead for cutting a single wire, where the qubit initialization depends on the previous measurement result via classical communication, has been determined to be $\kappa = 3$~\cite{Brenner2023}. 
Harada et al.~\cite{Harada2023} demonstrated a wire cut with minimal sampling overhead by providing the following decomposition of the single-qubit identity channel $\operatorname{Id}$:
\begin{align}
    \operatorname{Id}(\bullet)
    &=\sum_{i\in\{1,2\}} \sum_{j\in\{0,1\}} \mathrm{Tr}\left[ U_{i} \ket{j}\bra{j}U_{i}^{\dagger}(\bullet) \right]U_{i} \ket{j}\bra{j}U_{i}^{\dagger} - \sum_{j\in\{0,1\}} \mathrm{Tr}\left[ \ket{j}\bra{j}(\bullet) \right]X \ket{j}\bra{j}X,
\end{align}
where $U_1=H$ and $U_2=SH$, with $H$ the Hadamard gate and $S$ the phase gate.

\begin{figure}[t]
    \centering
    \input{figures/wire_cut_harada2}
    \caption{Optimal wire cut for a single wire~\cite{Harada2023}.}
    \label{fig:wire_cut_harada}
\end{figure}
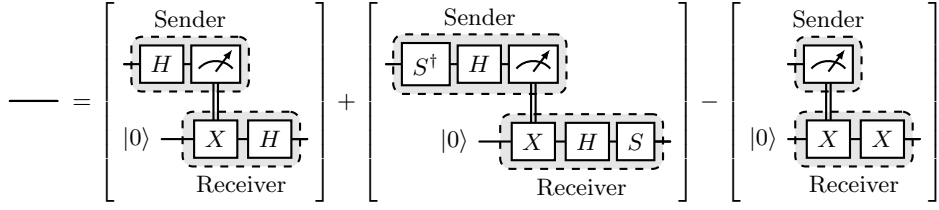

The wire cut circuits are shown in \Cref{fig:wire_cut_harada}, with measurements performed on a quantum device at the sender's side and the corresponding states initialized on a quantum device at the receiver's side.
Classical communication between the devices is facilitated by a classical controlled-not gate, represented by a double line connecting the measurement and the $X$ gate.

\subsection{Non-maximally entangled qubit pairs}
By making use of the Schmidt decomposition~\cite{Nielsen2009}, any entangled pure two-qubit state $\ket{\psi}$ can be expressed as 
\begin{align}\label{eq:schmidt}
    \ket{\psi} &= p_0 \ket{\xi_0} \ket{\zeta_0} + p_1 \ket{\xi_1} \ket{\zeta_1},
\end{align}
with $p_i \in \mathbb{R}_{\geq 0}$ and orthogonal sets of states $\{\zeta_i\}$ and $\{\xi_i\}$.
This representation shows that any two-qubit state can be reformulated as 
\begin{align}
    \ket{\psi} &= (A \otimes B) (\ket{\Phi^k}).
\end{align}
Herein  $A$ and $B$ are local unitary transformations that relate the Schmidt basis states to the computational basis states, and $\ket{\Phi^k}$ is a generalization of the Bell basis state $\ket{\Phi^+}$ given as 
\begin{align}\label{eq:nme}
    \ket{\Phi^{k}} = K(\ket{00} + k\ket{11})
\end{align}
with $K := \frac{1}{\sqrt{1+ k^{2}}}$ and $k \in \mathbb{R}_{\ge 0}$.
It is worth noting that the Schmidt decomposition of any bipartite pure state can be computed on near-term quantum devices, given a sufficient number of copies of the state~\cite{BravoPrieto2020}.
Therefore, in the following, we solely focus on NME pairs of qubits of the form in \Cref{eq:nme}.

To quantify the entanglement in the state $\ket{\Phi^k}$, we employ the \textit{robustness of entanglement}~\cite{Vidal1999}.
This entanglement measure determines how much noise can be added to a state before it becomes separable.
For a pure two-qubit state $\ket{\psi}$ with Schmidt coefficients $p_0$ and $p_1$ as defined in \Cref{eq:schmidt}, the robustness $R(\ket{\psi})$ is given by
\begin{align}
    R(\ket{\psi}) = (p_0 + p_1)^2 - 1.
\end{align}
The robustness of entanglement ranges between 0 and 1, with 0 denoting the absence of entanglement and 1 indicating a maximally entangled state.
The robustness of state $\ket{\Phi^k}$ is 
\begin{align}
    R(\ket{\Phi^k}) = 2kK^2.
\end{align}
Hence, the state $\ket{\Phi^k}$ is separable for $k=0$ and $k\to\infty$, while it is maximally entangled for $k=1$.

\subsection{Quantum Teleportation}
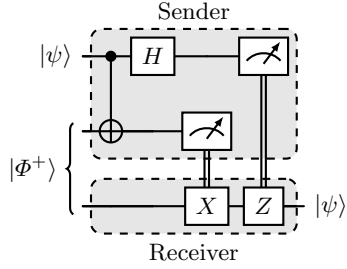
\begin{figure}[t]
    \centering
    \input{figures/teleportation_circuit}
    \caption{Teleportation circuit}
    \label{fig:teleportation_circuit}
\end{figure}

Similar to wire cutting, quantum teleportation enables measuring an unknown qubit state of a wire and initializing it on another wire~\cite{Bennett1993}.
This process involves the communication of only two classical bits without physically transmitting the state itself. 
To facilitate quantum teleportation, a shared pair of maximally entangled qubits, known as the resource state, is utilized between the sender and receiver. 
One example of such a resource state is $\ket{\Phi^{+}} = \frac{1}{\sqrt{2}}(\ket{00} + \ket{11})$.
The teleportation circuit, depicted in \Cref{fig:teleportation_circuit}, involves the sender performing a Bell basis measurement on the qubit to be teleported and his qubit of the entangled qubit pair. 
The measurement outcome is then transmitted to the receiver through a classical channel. 
Using this information, the receiver applies appropriate quantum operations to their part of the entangled pair, effectively reconstructing the original state at the receiver's location.
This is in contrast to wire cutting, where only the expectation value is reconstructed.
\looseness=-1

%% file: figures/wire_cut_harada2.tex
\[
    \begin{quantikz}[column sep=3pt, align equals at = 1]
    	\qw & \qw & \qw & \qw & \qw & \qw & \qw
    \end{quantikz}
    =
    \left[
    \rule{0cm}{1.45cm}
    \begin{quantikz}[column sep=3pt]
    	& \qw & \gate{H}\gategroup[1,steps=2,style={dashed,rounded corners,fill=gray!20, inner xsep=0pt, inner ysep=0pt},background, label style={label position=above,anchor=north,yshift=0.25cm}]{Sender} & \meter{}  \wire[d][1]{c}\\
        \setwiretype{n} & & \lstick{$\ket{0}$}  & \gate{X} \setwiretype{q} \gategroup[1,steps=2,style={dashed,rounded corners,fill=gray!20, inner xsep=0pt, inner ysep=0pt},background, label style={label position=below, anchor=north,yshift=-0.2cm}]{Receiver}& \gate{H} & \qw & \qw
    \end{quantikz}  
    \right]
    +
    \left[
    \rule{0cm}{1.45cm}
    \begin{quantikz}[column sep=3pt]
    	& \qw & \gate{S^\dagger}\gategroup[1,steps=3,style={dashed,rounded corners,fill=gray!20, inner xsep=0pt, inner ysep=0pt},background, label style={label position=above,anchor=north,yshift=0.25cm}]{Sender}  & \gate{H} & \meter{}  \wire[d][1]{c}\\
        \setwiretype{n} & & & \lstick{$\ket{0}$}  & \gate{X} \setwiretype{q} \gategroup[1,steps=3,style={dashed,rounded corners,fill=gray!20, inner xsep=0pt, inner ysep=0pt},background, label style={label position=below, anchor=north,yshift=-0.2cm}]{Receiver}& \gate{H} & \gate{S} & \qw & \qw
    \end{quantikz}
    \right]
    -
    \left[
    \rule{0cm}{1.45cm}
    \begin{quantikz}[column sep=3pt]
    	& \qw & \meter{}\gategroup[1,steps=1,style={dashed,rounded corners,fill=gray!20, inner xsep=0pt, inner ysep=0pt},background, label style={label position=above,anchor=north,yshift=0.25cm}]{Sender} \wire[d][1]{c}\\
         \lstick{$\ket{0}$} & & \gate{X} \setwiretype{q} \gategroup[1,steps=2,style={dashed,rounded corners,fill=gray!20, inner xsep=0pt, inner ysep=0pt},background, label style={label position=below, anchor=north,yshift=-0.2cm}]{Receiver}& \gate{X} & \qw & \qw
    \end{quantikz}  
    \right]
\]

%% file: figures/teleportation_circuit.tex
\[
    \begin{quantikz}[column sep=3pt, align equals at = 1]
    	\lstick{$\ket{\psi}$} & \qw  & \ctrl{1}\gategroup[2,steps=4,style={dashed,rounded corners,fill=gray!20, inner xsep=0pt, inner ysep=0pt},background, label style={label position=above,anchor=north,yshift=0.25cm}]{Sender} & \gate{H} & & \meter{} \wire[d][2]{c}  \\
    	\lstick[2]{$\ket{\Phi^{+}}$} & \qw & \targ{} & \qw & \meter{} \wire[d][1]{c} \\
    	& \qw & \qw \gategroup[1,steps=4,style={dashed,rounded corners,fill=gray!20, inner xsep=0pt, inner ysep=0pt},background, label style={label position=below, anchor=north,yshift=-0.2cm}]{Receiver} & \qw & \gate{X} & \gate{Z} & \qw &\rstick{$\ket{\psi}$} 
    \end{quantikz}
\]

%% file: content/3_main_result.tex
\section{Approach}\label{sec:main-section}
Our objective is to enhance the wire cutting procedure by utilizing NME qubit pairs $\ket{\Phi^{k}}$  with different degrees of entanglement as given by the parameter $k$.
This allows for reducing the sampling overhead for the wire cutting procedure by increasing the degree of entanglement. 
Furthermore, by adjusting the degree of entanglement, this approach shows the relationship between wire cutting and the quantum teleportation protocol: for minimal entanglement (i.e., $k=0$), the presented approach reduces to wire cutting, and for maximal entanglement (i.e., $k=1$) the approach is equivalent to the quantum teleportation protocol.

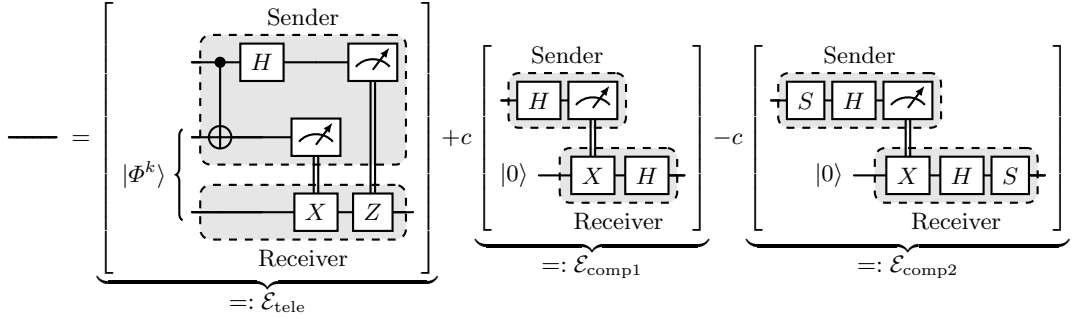
\begin{figure}[t]
    \centering
    \input{figures/cut_with_entanglement3}
    \caption{Wire cut that leverages non-maximally entangled states.}
    \label{fig:cut_with_entanglement}
\end{figure}

On a high-level view, the decomposition of the wire, as depicted in \Cref{fig:cut_with_entanglement}, consists of two components: (i)~the quantum teleportation circuit ($\mathcal{E}_{\text{tele}}$) that deviates from the conventional teleportation protocol by employing an NME resource state $\ket{\Phi^{k}}$ and (ii)~two circuits ($\mathcal{E}_{\text{comp1}}$ and $\mathcal{E}_{\text{comp2}}$) that are weighted by a factor $c \in [0,1]$, which depends on the degree of entanglement in the resource state $\ket{\Phi^{k}}$. 
This factor $c$ compensates for the deficiency of the teleportation protocol when using NME states, and therefore we call these two circuits \textit{compensation circuits}.
When $k=1$ and therefore $\ket{\Phi^{k}}$ is maximally entangled, the teleportation circuit faithfully transmits the exact state without requiring compensation, resulting in $c=0$ and no sampling overhead since no shots have to be sampled from the compensation circuits.
Conversely, if $k=0$, the state $\ket{\Phi^{k}}$ lacks any entanglement and a default wire cutting procedure is employed, yielding $c=1$ and a sampling overhead characterized by $\kappa = 3$.

On the input of a general one-qubit state $\ket{\psi} = \alpha\ket{0} + \beta\ket{1}$, the teleportation protocol with $\ket{\Phi^{k}}$ as resource state yields the following mixed state:
\begin{align}
    \mathcal{E}_{\text{tele}}(\ketbras{\psi}) &= |\alpha|^2\ketbras{0} + |\beta|^2\ketbras{1} + 2kK^{2}\alpha\beta^{*}\ketbra{0}{1} +  2kK^{2}\alpha^{*}\beta \ketbra{1}{0}.
\end{align}
The resulting density operator deviates from $\ketbras{\psi}$ in the magnitude of its off-diagonal elements $\ketbra{0}{1}$ and $\ketbra{1}{0}$, which are scaled by the robustness value $R(\ket{\Phi^k})$. 
This deviation occurs due to the use of the NME resource state $\ket{\Phi^{k}}$ in the teleportation protocol.
The compensation circuits described above are used to account for the scaled off-diagonal elements. 
For the same input, the difference between the two compensation circuits yields
\begin{align}
    c \mathcal{E}_{\text{comp1}}(\ketbras{\psi}) - c \mathcal{E}_{\text{comp2}}(\ketbras{\psi}) = c \alpha\beta^{*}\ketbra{0}{1} + c \alpha^{*}\beta\ketbra{1}{0}.
\end{align}
Therefore, by applying the compensation circuits with a compensation factor $c = 1 - R(\ket{\Phi^k})$, which accounts for the lack of entanglement indicated by the robustness of the resource state,  it is possible to recover the original state:
\begin{align}
    \mathcal{E}_{\text{tele}}(\ketbras{\psi}) + \big(1 - R(\ket{\Phi^k})\big)\big(\mathcal{E}_{\text{comp1}}(\ketbras{\psi}) - \mathcal{E}_{\text{comp2}}(\ketbras{\psi})\big) = \ketbras{\psi}.
\end{align}
Consequently, the factor $\kappa_{\text{NME}}(k)$ that characterizes the sampling overhead for wire cutting with NME state $\ket{\Phi^k}$ is given by
\begin{align}
    \kappa_{\text{NME}}(k) = 3 - 2R(\ket{\Phi^k}) = 3 - \frac{4k}{1+k^2}.
\end{align}
It is lower bounded by $1$ when using a maximally entangled state ($k=1$), and it is upper bounded by $3$ when employing separable resource states ($k=0$ and $k \to \infty$).

%% file: figures/cut_with_entanglement3.tex
\[
    \begin{quantikz}[column sep=3pt, align equals at = 1]
    	\qw & \qw & \qw & \qw & \qw & \qw & \qw
    \end{quantikz}
    =
    \underbrace{
    \left[
    \rule{0cm}{1cm}
    \begin{quantikz}[column sep=3pt, align equals at = 2]
    	 & \qw & \ctrl{1}\gategroup[2,steps=4,style={dashed,rounded corners,fill=gray!20, inner xsep=0pt, inner ysep=0pt},background, label style={label position=above,anchor=north,yshift=0.25cm}]{Sender} & \gate{H} & & \meter{} \wire[d][2]{c}  \\
    	\lstick[2]{$\ket{\Phi^k}$} & \qw  & \targ{} & \qw & \meter{} \wire[d][1]{c} \\
    	& \qw & \qw\gategroup[1,steps=4,style={dashed,rounded corners,fill=gray!20, inner xsep=0pt, inner ysep=0pt},background, label style={label position=below, anchor=north,yshift=-0.2cm}]{Receiver} & \qw & \gate{X} & \gate{Z} & \qw &
    \end{quantikz}
    \right]}_{\textstyle =:\mathcal{E}_{\text{tele}}}
    + c
    \underbrace{
    \left[
    \rule{0cm}{1cm}
    \begin{quantikz}[column sep=3pt]
    	& \qw & \gate{H}\gategroup[1,steps=2,style={dashed,rounded corners,fill=gray!20, inner xsep=0pt, inner ysep=0pt},background, label style={label position=above,anchor=north,yshift=0.25cm}]{Sender} & \meter{}  \wire[d][1]{c}\\
        \setwiretype{n} & & \lstick{$\ket{0}$}  & \gate{X} \setwiretype{q} \gategroup[1,steps=2,style={dashed,rounded corners,fill=gray!20, inner xsep=0pt, inner ysep=0pt},background, label style={label position=below, anchor=north,yshift=-0.2cm}]{Receiver}& \gate{H} & \qw & \qw
    \end{quantikz}
    \right]}_{\textstyle =:\mathcal{E}_{\text{comp1}}}
    - c
    \underbrace{
    \left[
    \begin{quantikz}[column sep=3pt]
    	& \qw & \gate{S}\gategroup[1,steps=3,style={dashed,rounded corners,fill=gray!20, inner xsep=0pt, inner ysep=0pt},background, label style={label position=above,anchor=north,yshift=0.25cm}]{Sender}  & \gate{H} & \meter{}  \wire[d][1]{c}\\
        \setwiretype{n} & & & \lstick{$\ket{0}$}  & \gate{X} \setwiretype{q} \gategroup[1,steps=3,style={dashed,rounded corners,fill=gray!20, inner xsep=0pt, inner ysep=0pt},background, label style={label position=below, anchor=north,yshift=-0.2cm}]{Receiver}& \gate{H} & \gate{S} & \qw & \qw
    \end{quantikz}
    \right]}_{\textstyle =:\mathcal{E}_{\text{comp2}}}
\]

%% file: content/4_numerical_experiments.tex
\section{Numerical Experiments}\label{sec:num_experiments}

This section presents a numerical demonstration showcasing the advantages of incorporating NME states in the wire cutting procedure.
Our experimental setup involves applying a wire cut to a qubit initialized in a random state and assessing the resulting error for varying numbers of shots.
For each random state, the cutting procedure is conducted with NME states according to \Cref{eq:nme}, with varying robustness values of entanglement.
We follow a specific procedure to initialize a qubit in a random state: 
A unitary matrix $U$ is randomly sampled~\cite{Mezzadri2007} and applied to the initial state $\ket{0}$. 
The resulting state is given by $U\!\ket{0}$, and the exact measurement probabilities of $\ket{0}$ and $\ket{1}$ can be computed as $|\!\bra{0}\!U\!\ket{0}\!|^2$ and $|\!\bra{1}\!U\!\ket{0}\!|^2$, respectively.

\begin{figure}[t]
    \centering
    \includegraphics[width=0.6\linewidth]{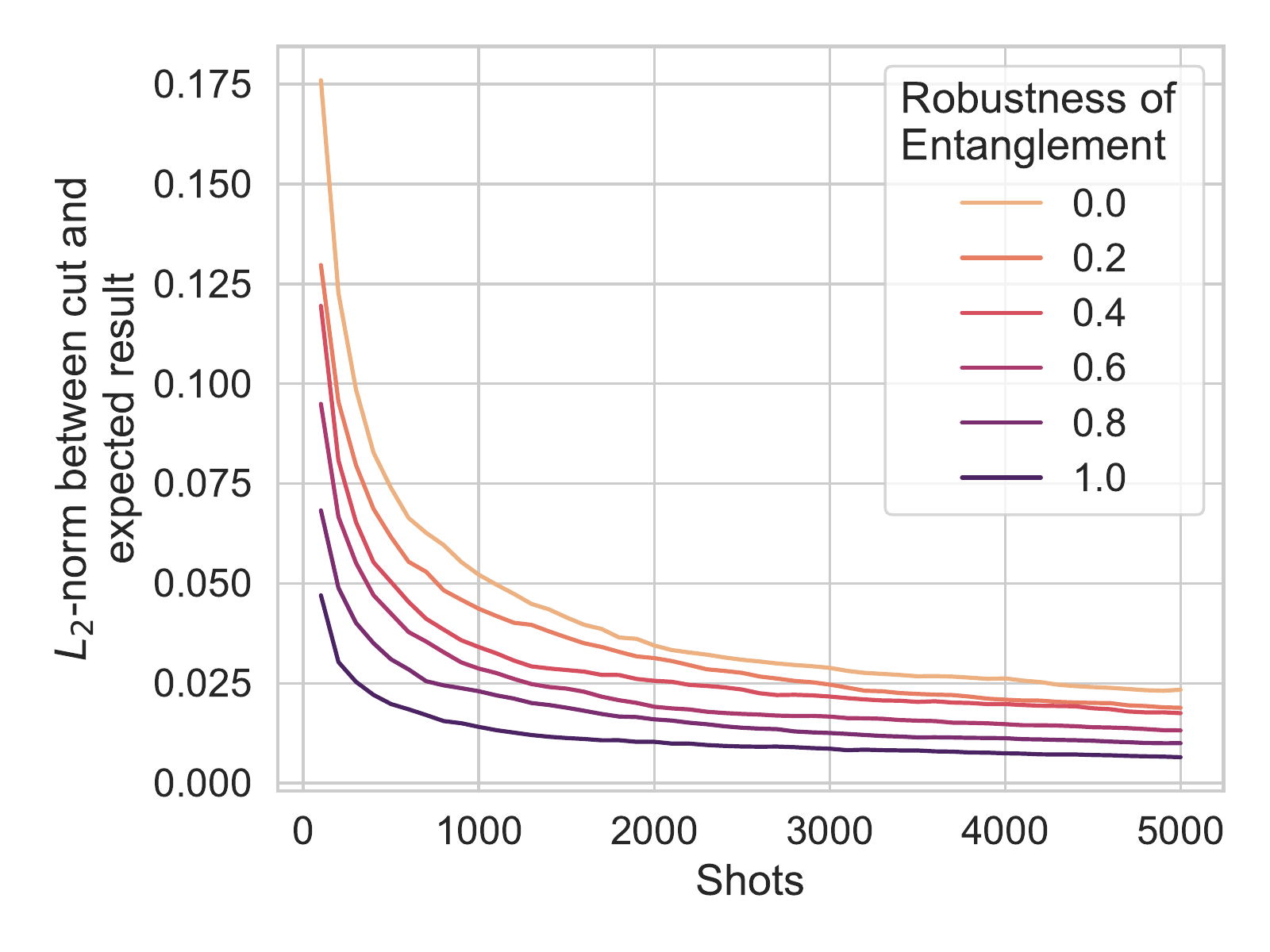}
    \caption{Average $L_{2}$-norm between the cut and expected result for resource states with different robustness of entanglement.}
    \label{fig:robustness_plot} 
\end{figure}
Next, the measurement probabilities with wire cutting are computed.
The wire cut is applied to the state $U\!\ket{0}$, resulting in three subcircuits: as described above one teleportation circuit and two compensation circuits. 
The resulting state of each subcircuit is measured in the computational basis. 
To execute these subcircuits, a simulator is utilized. 
A fixed number of shots is allocated collectively to all three subcircuits, distributed proportionally to their probabilities as defined in \Cref{eq:expectation}.
The deviation between the simulated result with wire cutting and the exact measurement probabilities, considering the number of shots and the robustness of entanglement of the NME state employed in the cut, indicates the sampling overhead.
For this evaluation, we considered 500 random states and performed cuts on them using NME states characterized by robustness values of 0, 0.2, 0.4, 0.6, 0.8, and 1. 

The presented results in \Cref{fig:robustness_plot} demonstrate a clear relationship between the robustness of entanglement in the resource states and the deviation from the exact result, measured in terms of the $L_{2}$-norm between the cut and exact results.
Higher robustness values of entanglement correspond to lower deviations from the exact result. 
Additionally, we observe that using more highly entangled states leads to faster convergence in the number of shots.
The case where the robustness is 1 serves as a baseline, employing default quantum teleportation.
In this scenario, no shots are allocated to the compensation circuits, and the deviation from the exact result arises only from statistical errors caused by the finite number of shots since the exact state is teleported.
Conversely, when the robustness is 0, representing default wire cutting, the $L_{2}$-norm between the cut and the exact result is the largest.
These observations align with the theoretical considerations regarding the sampling overhead of wire cutting with NME states in Section \ref{sec:main-section}.


%% file: content/5_related_work.tex
\section{Related Work}\label{seq:related_work}
Piveteau and Sutter~\cite{Piveteau2022} showed a  one-to-one relationship between the optimal sampling overhead characterized by $\kappa_{\rho}$ necessary for simulating a bipartite state $\rho$ using a quasi-probabilistic decomposition limited to local operations acting on a fixed product state, and the state's robustness of entanglement $R(\rho)$. 
Specifically, they found that $\kappa_{\rho} = 1 + 2R(\rho)$.
Our findings align with this result, as the sampling overhead of our proposed wire cutting procedure coincides with the overhead obtained when simulating the bipartite state $\ket{\Phi^+}$ using a quasi-probabilistic decomposition involving local operations on $\ket{\Phi^k}$.
The simulated state $\ket{\Phi^+}$ allows for faithful teleportation and, thus, can be utilized for wire cutting.
Consequently, the sampling overhead of our procedure, characterized by $\kappa_{\text{NME}}(k)$, effectively reflects the overhead that accounts for the difference in robustness of entanglement between the states $\ket{\Phi^+}$ and $\ket{\Phi^k}$, which is given by
\begin{align}
    1 + 2\left(R(\ket{\Phi^+}) - R(\ket{\Phi^k})\right) = 3 - 2 R(\ket{\Phi^k}) = \kappa_{\text{NME}}(k).
\end{align}

Moreover, an alternative circuit cutting technique to wire cutting is gate cutting~\cite{Mitarai2021,Piveteau2022}. 
It involves decomposing multi-qubit gates to cut a circuit. 
Depending on the characteristics of the circuit, either a wire cut or gate cut can be more favorable to achieve a minimal sampling overhead~\cite{Brenner2023}. 
Additionally, it has been demonstrated that cutting multiple wires together can effectively reduce the sampling overhead, compared to cutting each one individually~\cite{Brenner2023}.
A framework for automatically finding optimal positions for wire cuts has been developed~\cite{Tang2021}.
Moreover, to address finite-shot error, maximum likelihood methods have been introduced in wire cutting~\cite{Perlin2021}.
Although circuit cutting primarily aims to reduce circuit size, some studies empirically demonstrate that executing smaller subcircuits can enhance overall results~\cite{Ayral2021,Bechtold2023,Perlin2021}.

Additionally, previous research has investigated using NME states as resource states in quantum teleportation. 
When employing the original protocol with NME states, the fidelity of the teleported state is reduced~\cite{Prakash2012}. 
Unit fidelity can only be achieved when using maximally entangled states. 
However, the modified probabilistic teleportation protocol allows the teleportation of an unknown state with NME resource states while maintaining unit fidelity~\cite{Agrawal2002,Pati2004}. 
Nevertheless, the success probability of this protocol is less than unity, requiring the repetition of the protocol in the event of failure and introducing an overhead to achieve the desired number of successful teleportations.

%% file: content/6_discussion_conclusion.tex
\section{Conclusion}\label{sec:conclusion}
The wire cutting technique proposed in this work takes advantage of pure NME states to lower its associated cost in terms of the sampling overhead. 
By increasing the degree of entanglement in the resource states, the sampling overhead can be effectively reduced. 
This highlights the significance of entanglement as a valuable computational resource. 
Furthermore, our approach allows for interpolation between an optimal wire cut without entanglement and quantum teleportation employing maximally entangled states.
While this paper is written from the perspective of enhancing circuit cutting with entanglement, another perspective is that of applying error mitigation to the teleportation protocol that accounts for missing entanglement in the resource state.
With the findings presented in this paper, we can now bridge the gap between circuit cutting and quantum teleportation using NME resource states, bringing us closer to the realization of distributed quantum computing.


Future work can explore how the results of the utilization of NME states in wire cutting transfers to gate cutting techniques, enhancing their efficiency in decomposing multi-qubit gates.
Additionally, investigating the application of NME states for multiple cuts in parallel could offer further improvements in reducing the sampling overhead. 
Another interesting avenue for research is the exploration of NME mixed states and their potential benefits in circuit cutting procedures.

%% file: main.bbl
\begin{thebibliography}{10}
\providecommand{\url}[1]{\texttt{#1}}
\providecommand{\urlprefix}{URL }
\providecommand{\doi}[1]{https://doi.org/#1}

\bibitem{Agrawal2002}
Agrawal, P., Pati, A.K.: {Probabilistic Quantum Teleportation}. Physics Letters
  A  \textbf{305}(1-2),  12--17 (2002)

\bibitem{Avron2021}
Avron, J., Casper, O., Rozen, I.: Quantum advantage and noise reduction in
  distributed quantum computing. Physical Review A  \textbf{104}(5),  052404
  (2021)

\bibitem{Ayral2021}
Ayral, T., Régent, F.M.L., Saleem, Z., Alexeev, Y., Suchara, M.: Quantum
  divide and compute: Exploring the effect of different noise sources. SN
  Computer Science  \textbf{2}(3), ~132 (2021)

\bibitem{Bechtold2023}
Bechtold, M., Barzen, J., Leymann, F., Mandl, A., Obst, J., Truger, F., Weder,
  B.: {Investigating the effect of circuit cutting in QAOA for the MaxCut
  problem on NISQ devices}  (2023), arXiv:2302.01792

\bibitem{Bennett1993}
Bennett, C.H., Brassard, G., Cr{\'{e}}peau, C., Jozsa, R., Peres, A., Wootters,
  W.K.: {Teleporting an unknown quantum state via dual classical and
  Einstein-Podolsky-Rosen channels}. Physical Review Letters  \textbf{70}(13),
  1895--1899 (1993)

\bibitem{BravoPrieto2020}
Bravo-Prieto, C., García-Martín, D., Latorre, J.I.: Quantum singular value
  decomposer. Physical Review A  \textbf{101}(6),  062310 (2020)

\bibitem{Bravyi2022}
Bravyi, S., Dial, O., Gambetta, J.M., Gil, D., Nazario, Z.: The future of
  quantum computing with superconducting qubits. Journal of Applied Physics
  \textbf{132}(16),  160902 (2022)

\bibitem{Bravyi2016}
Bravyi, S., Smith, G., Smolin, J.A.: Trading {Classical} and {Quantum}
  {Computational} {Resources}. Physical Review X  \textbf{6}(2),  021043 (2016)

\bibitem{Brenner2023}
Brenner, L., Piveteau, C., Sutter, D.: Optimal wire cutting with classical
  communication  (2023), arXiv:2302.03366

\bibitem{Cao2018}
Cao, Y., Romero, J., Aspuru-Guzik, A.: Potential of quantum computing for drug
  discovery. {IBM} Journal of Research and Development  \textbf{62}(6),
  6:1--6:20 (2018)

\bibitem{Cao2019}
Cao, Y., et~al.: Quantum {Chemistry} in the {Age} of {Quantum} {Computing}.
  Chemical Reviews  \textbf{119}(19),  10856--10915 (2019)

\bibitem{Cuomo2020}
Cuomo, D., Caleffi, M., Cacciapuoti, A.S.: Towards a distributed quantum
  computing ecosystem. {IET} Quantum Communication  \textbf{1}(1), ~3--8 (2020)

\bibitem{Dunjko2018}
Dunjko, V., Ge, Y., Cirac, J.I.: Computational speedups using small quantum
  devices. Physical Review Letters 121, 250501  (2018)

\bibitem{Furutanpey2023_QuantumEdge}
Furutanpey, A., Barzen, J., Bechtold, M., Dustdar, S., Leymann, F., Raith, P.,
  Truger, F.: {Architectural Vision for Quantum Computing in the Edge-Cloud
  Continuum}  (2023), arXiv:2305.05238

\bibitem{Harada2023}
Harada, H., Wada, K., Yamamoto, N.: Optimal parallel wire cutting without
  ancilla qubits  (2023), arXiv:2303.07340

\bibitem{Khait2023}
Khait, I., Tham, E., Segal, D., Brodutch, A.: {Variational Quantum Eigensolvers
  in the Era of Distributed Quantum Computers}  (2023), arXiv:2302.03366

\bibitem{Mezzadri2007}
Mezzadri, F.: How to generate random matrices from the classical compact
  groups. Notices of the American Mathematical Society  \textbf{54}(5),  592 --
  604 (2007)

\bibitem{Mitarai2021}
Mitarai, K., Fujii, K.: Constructing a virtual two-qubit gate by sampling
  single-qubit operations. New Journal of Physics  \textbf{23}(2),  023021
  (2021)

\bibitem{Nielsen2009}
Nielsen, M.A., Chuang, I.L.: Quantum Computation and Quantum Information.
  Cambridge University Press (2009)

\bibitem{Pati2004}
Pati, A.K., Agrawal, P.: Probabilistic teleportation and quantum operation.
  Journal of Optics B: Quantum and Semiclassical Optics  \textbf{6}(8),
  S844--S848 (2004)

\bibitem{Pednault2023}
Pednault, E.: An alternative approach to optimal wire cutting without ancilla
  qubits  (2023), arXiv:2303.08287

\bibitem{Peng2019}
Peng, T., Harrow, A., Ozols, M., Wu, X.: {Simulating Large Quantum Circuits on
  a Small Quantum Computer}. Physical Review Letters  \textbf{125},  150504
  (2019)

\bibitem{Perlin2021}
Perlin, M.A., Saleem, Z.H., Suchara, M., Osborn, J.C.: Quantum circuit cutting
  with maximum likelihood tomography. npj Quantum Information  \textbf{7}(1)
  (2021)

\bibitem{Piveteau2022}
Piveteau, C., Sutter, D.: Circuit knitting with classical communication
  (2022), arXiv:2205.00016

\bibitem{Prakash2012}
Prakash, H., Verma, V.: Minimum assured fidelity and minimum average fidelity
  in quantum teleportation of single qubit using non-maximally entangled
  states. Quantum Information Processing  \textbf{11}(6),  1951--1959 (2012)

\bibitem{Preskill2018}
Preskill, J.: {Quantum Computing in the NISQ era and beyond}. Quantum 2, 79
  (2018)  \textbf{2}, ~79 (2018)

\bibitem{Tang2021}
Tang, W., Tomesh, T., Suchara, M., Larson, J., Martonosi, M.: {CutQC: using
  small Quantum computers for large Quantum circuit evaluations}. In:
  Proceedings of the 26th {ACM} International Conference on Architectural
  Support for Programming Languages and Operating Systems. pp. 473--486. ACM
  (2021)

\bibitem{Temme2017}
Temme, K., Bravyi, S., Gambetta, J.M.: {Error Mitigation for Short-Depth
  Quantum Circuits}. Physical Review Letters  \textbf{119}(18),  180509 (2017)

\bibitem{Vidal1999}
Vidal, G., Tarrach, R.: Robustness of entanglement. Physical Review A
  \textbf{59}(1),  141--155 (1999)

\end{thebibliography}
